\documentclass[aps,reprint,notitlepage,floatfix,superscriptaddress]{revtex4-1}

\usepackage[pdftex]{graphicx}
\usepackage{amsmath,amssymb,amstext}
\usepackage{mathrsfs}
\usepackage{multirow}
\usepackage{comment}
\usepackage{bbold}
\usepackage{dsfont}
\usepackage{color}
\usepackage{hyperref}
\usepackage{tikz}
\usepackage{braket}
\usepackage[colorinlistoftodos]{todonotes}
\usepackage{soul}
\renewcommand{\eqref}[1]{Eq.~(\ref{#1})}
\newcommand{\figref}[1]{Fig.~\ref{#1}}

\begin{document}
\title{Remote state preparation of single photon orbital angular momentum lattices}

\author{Andrew R. Cameron}
\email{ar3camer@uwaterloo.ca}
\affiliation{Institute for Quantum Computing, University of Waterloo, Waterloo, ON N2L 3G1, Canada}
\affiliation{Department of Physics \& Astronomy, University of Waterloo,Waterloo, ON N2L 3G1, Canada}

\author{Sandra W. L. Cheng}
\email{sandra.cheng@uwaterloo.ca}
\affiliation{Institute for Quantum Computing, University of Waterloo, Waterloo, ON N2L 3G1, Canada}
\affiliation{Department of Physics \& Astronomy, University of Waterloo,Waterloo, ON N2L 3G1, Canada}

\author{Sacha Schwarz}
\affiliation{Institute for Quantum Computing, University of Waterloo, Waterloo, ON N2L 3G1, Canada}
\affiliation{Department of Physics \& Astronomy, University of Waterloo,Waterloo, ON N2L 3G1, Canada}
\affiliation{Infinite Potential Laboratories LP, 485 Wes Graham Way, Waterloo, ON N2L A07, Canada}

\author{Connor Kapahi}
\affiliation{Institute for Quantum Computing, University of Waterloo, Waterloo, ON N2L 3G1, Canada}
\affiliation{Department of Physics \& Astronomy, University of Waterloo,Waterloo, ON N2L 3G1, Canada}

\author{Dusan Sarenac}
\affiliation{Institute for Quantum Computing, University of Waterloo, Waterloo, ON N2L 3G1, Canada}

\author{Michael Grabowecky}
\affiliation{Institute for Quantum Computing, University of Waterloo, Waterloo, ON N2L 3G1, Canada}
\affiliation{Department of Physics \& Astronomy, University of Waterloo,Waterloo, ON N2L 3G1, Canada}

\author{David G. Cory}
\affiliation{Institute for Quantum Computing, University of Waterloo, Waterloo, ON N2L 3G1, Canada}
\affiliation{Department of Chemistry, University of Waterloo, Waterloo, ON N2L 3G1, Canada}

\author{Thomas Jennewein}
\affiliation{Institute for Quantum Computing, University of Waterloo, Waterloo, ON N2L 3G1, Canada}
\affiliation{Department of Physics \& Astronomy, University of Waterloo,Waterloo, ON N2L 3G1, Canada}

\author{Dmitry A. Pushin}
\affiliation{Institute for Quantum Computing, University of Waterloo, Waterloo, ON N2L 3G1, Canada}
\affiliation{Department of Physics \& Astronomy, University of Waterloo,Waterloo, ON N2L 3G1, Canada}

\author{Kevin J. Resch}
\affiliation{Institute for Quantum Computing, University of Waterloo, Waterloo, ON N2L 3G1, Canada}
\affiliation{Department of Physics \& Astronomy, University of Waterloo,Waterloo, ON N2L 3G1, Canada}

\begin{abstract}

Optical beams with periodic lattice structures have broadened the study of structured waves. In the present work, we generate spin-orbit entangled photon states with a lattice structure and use them in a remote state preparation protocol. We sequentially measure spatially-dependent correlation rates with an electron-multiplying intensified CCD camera and verify the successful remote preparation of spin-orbit states by performing pixel-wise quantum state tomography. Control of these novel structured waves in the quantum regime provides a method for quantum sensing and manipulation of periodic structures.

\end{abstract}

\maketitle

\section{Introduction}

Advances in experimental methods have enabled the creation of structured beams of neutrons~\cite{Sarenac20328}. Matter-waves with structured phase fronts are formed with many different strategies ranging from spatially-dependent magnetic fields~\cite{matter1,sarenac2018methods,Rubinsztein_Dunlop_2016} to spiral phase plates made of thin graphite films~\cite{matter2}. The formalism of quantum information science is system agnostic, allowing translation of the physics of one system to that of another. In order to move from neutrons to photons, spin is replaced with polarization, and the magnetic field gradients are replaced with birefringent gradients. Using this correspondence, a lattice of spin-orbit states originally developed for neutrons has been implemented with photons. Optical lattices have led to studies of optical Talbot physics of structured orbital angular momentum (OAM) light beams~\cite{talbot2,PhysRevA.101.043815}, optical lattice structure shaping~\cite{optical_lattice_shaping1,optical_lattice_shaping2}, and direct detection of optical spin-orbit states by the human eye~\cite{Sarenac14682,sarenac2020human}. By translating the physics of a periodic structure of spin-orbit states further in photonics, we can take advantage of additional capabilities such as multi-particle entanglement. This opens the possibility for \emph{quantum} correlations in structured beams and the capabilities that come with them. 

 The periodicity of these structured waves are suited for quantum sensing or control of periodic structures~\cite{Andersen2006,he1995direct,Schmiegelow2012}. The interference of OAM lattices has been used to build all-optical quantum memory devices~\cite{luo_synthetic-lattice_2017}, and the average deviation of atoms relative to their lattice sites has been measured continuously and nondestructively with optical lattices~\cite{optical_lattice}. OAM provides access to a high-dimensional Hilbert space which can enhance the information capacity of a single particle~\cite{Forbes2019,Barreiro2008}, while the more easily manipulated polarization degree of freedom can be used for enhanced control and measurement~\cite{Marrucci2011,Milione2015,Vallone2014,Schmiegelow2016,erhard2018twisted,fickler2016,Sit2017}. Working with the OAM and polarization degrees of freedom simultaneously combines the advantageous characteristics of both~\cite{Nagali2009,Wang2011,diamanti_practical_2016,Mafu2013,Heo2017}. To characterize and verify spin-orbit entanglement, quantum state tomography has been done previously using OAM projective measurements with a spatial light modulator~\cite{PhysRevA.92.022321}, and with spatially-dependent polarization measurements using an intensified CCD camera~\cite{PhysRevA.89.060301}. Structured waves have recently attracted attention in the quantum communication community specifically in turbulence studies~\cite{turbulence1,turbulence3}. Correlations between polarization and OAM have shown preservation of the encoded state after propagation through scattering media~\cite{turbulence2}.
 
In this work, we generate spin-orbit entanglement between the polarization of one photon and the transverse beam profile of the other. Polarization measurement enables production of distinctly different structured beams, and the correlations between these beams and the polarization can be used to verify entanglement. We confirm the entanglement using a quantum state tomography procedure between the polarization of one photon, and the position-dependent polarization of its entangled partner. With these correlations, we implement a remote state preparation (RSP) protocol to prepare structured single photon beams. RSP involves transferring a quantum state known by one party to another party via entanglement~\cite{Pati2000,Bennett2001,Lo2000}, and has applications in large-scale quantum communication networks~\cite{Leung2003,Dakic2012,barreiro2010remote,Peters2005}. In our case, a RSP protocol is used to prepare signal photon spatial patterns conditioned on idler photon polarization measurements. The spin-orbit coupling method presented expands lattice structured light preparation and measurement further into the quantum regime.


\section{Theory}

\begin{figure*}[ht!]
\centering
\includegraphics[trim={0 0 0 0},width=\textwidth]{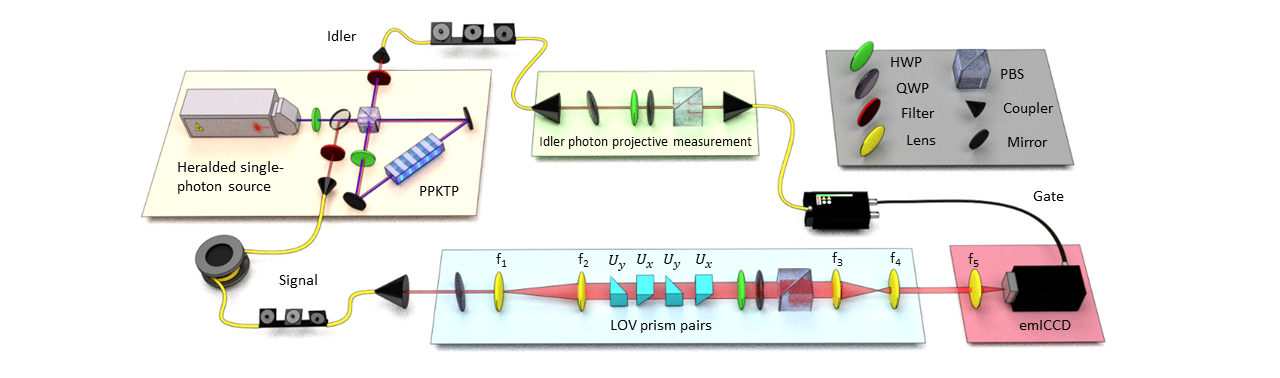}
\caption{Schematic of the experiment. Polarization-entangled photon pairs are generated via type-II spontaneous parametric down-conversion in a Sagnac interferometer and coupled into single mode fibers. After propagating through a $30$~m fiber, the signal photon is sent through a telescope with $8.3\times$ magnification ($f_1$ and $f_2$ lenses), two sets of ``Lattice of Optical Vortices'' (LOV) prism pairs and polarization analyzing optics. The signal photons are then gated to an electron-multiplying intensified CCD (emICCD) camera, triggered by the detection of the corresponding polarization-filtered idler. The imaging arrangement in the detection unit consists of a telescope with $4\times$ demagnification ($f_3$ and $f_4$ lenses) followed by a lens ($f_5$) that images the beam onto the detection plane of the emICCD camera.}
\label{fig:expSetup}
\end{figure*}

We consider polarization-entangled photon pairs which are described by the Bell state $\vert \Phi^+ \rangle = \frac{1}{\sqrt{2}}\left( \vert L R \rangle + \vert R L \rangle \right)$, where we denote right-handed circular and left-handed circular polarization states by $\vert R \rangle = \frac{1}{\sqrt{2}}(\vert H \rangle + i \vert V \rangle)$ and $\vert L \rangle = \frac{1}{\sqrt{2}}(\vert H \rangle - i \vert V \rangle)$. Polarization states $\vert H \rangle$ and $\vert V \rangle$ correspond to $\vert 0 \rangle$ and $\vert 1 \rangle$, respectively, in the computational basis. As described in Ref.~\cite{sarenac2018generation} in more detail, a lattice of spin-orbit states is obtained by passing circularly polarized light through perpendicular pairs of birefringent linear gradients whose optical axes are relatively offset by $45^{\circ}$. The operators of the two perpendicular birefringent gradients are described by
\begin{equation} 
    \hat{U}_x=e^{i\frac{\pi}{a}(x-x_0)\hat{\sigma}_x},
    \quad
    \hat{U}_y=e^{i\frac{\pi}{a}(y-y_0)\hat{\sigma}_z},
    \label{eq:shiftOperators}
\end{equation}
where the origin of the gradients is given by $(x_0, y_0)$, $\hat{\sigma}_{x,z}$ are Pauli matrices and $a=\lambda(\Delta n \tan(\theta))^{-1}$ is the spacing between neighboring lattice sites with wavelength $\lambda$, prism birefringence $\Delta n$ and prism incline angle $\theta$. By sending one photon through $N=2$ sets of \emph{Lattice of Optical Vortices} (LOV) prism pairs, we prepare the spin-orbit entangled lattice state
\begin{equation}
    \vert \Psi_\text{LOV}^{N=2} \rangle(x,y) = \frac{\alpha(x,y)}{\sqrt{2}}\left[(\hat{U}_x\hat{U}_y)^2 \otimes \mathbb{I}_2\right]\vert \Phi^+\rangle,
    \label{eq:LOVState}
\end{equation}
where $\alpha(x, y)$ describes the incoming Gaussian beam envelope, and $\mathbb{I}_2$ is the two-dimensional identity matrix. 

Applying the operators in \eqref{eq:shiftOperators} on polarization states $\vert L \rangle$ and $\vert R \rangle$  yields
\begin{equation}
    (\hat{U}_x\hat{U}_y)^2\vert L \rangle = A(x,y)\vert L \rangle + B(x,y)\vert R \rangle
    \label{eq:photonModulationL}
\end{equation}
and
\begin{equation}
    (\hat{U}_x\hat{U}_y)^2\vert R \rangle = A(x,y)\vert R \rangle + B(x,y)\vert L \rangle,
    \label{eq:photonModulationR}
\end{equation}
where $A(x,y)$ and $B(x,y)$ are complex-valued amplitudes. The LOV prism pairs are thus represented by unitary matrices that couple the polarization of a photon to its spatial mode. Different polarization projections on the spin-orbit lattice state lead to different intensity patterns. To simulate these intensity patterns, polarization projections were applied to \eqref{eq:LOVState}. Furthermore, we applied a Gaussian beam profile to the theoretical images in order to account for the beam intensity envelope. 

The two-photon density matrix is recovered via maximum likelihood quantum state tomography. The information of interest is encoded in the complex two-dimensional spatial functions as seen in \eqref{eq:photonModulationL} and \eqref{eq:photonModulationR}, and a single photon camera captures intensity measurements of the entire pattern simultaneously. Each of the camera's pixels are treated like individual detectors when computing tomography. A pixel-wise algorithm loops through them and uses the maximum likelihood tomography approach specified in Ref.~\cite{James2001}. By recovering the density matrix at every pixel, we can witness entanglement between polarization and each transverse position in the beam and verify remote state preparation. 

\section{Experimental Method}

\begin{figure*}[ht!]
\centering
\includegraphics[trim={5cm 0 0 3cm},width=\textwidth]{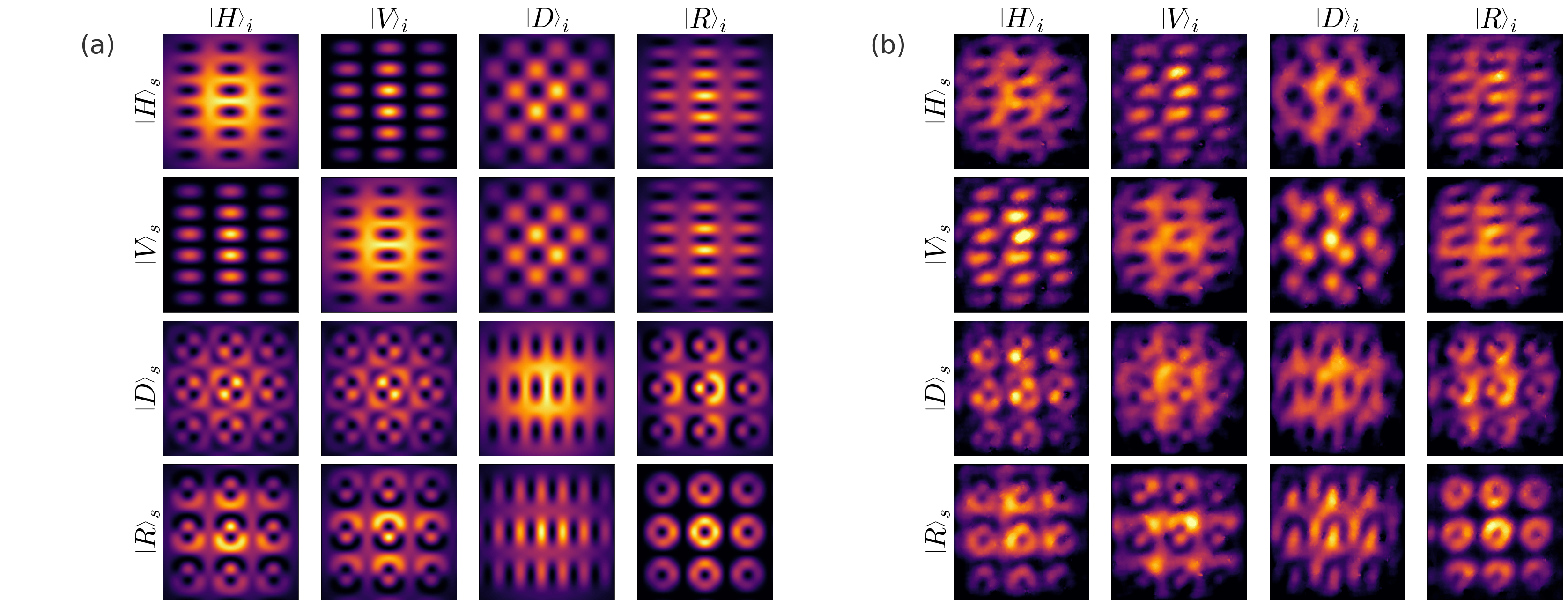}
\caption{Theoretical predictions and experimental results. Rows (columns) are organized by signal (idler) projective measurements labeled by the corresponding polarization. (a) Theoretical evaluation of the 16 measurement configurations over a $140 \times 140$ grid. (b) Experimental results of the 16 measurement configurations over a $140 \times 140$ pixel area. All theoretical intensity patterns are normalized and all experimental intensity patterns are normalized and post-processed using background subtraction and an adaptive two-dimensional Gaussian image filter. The emICCD camera records photon counts, and an artificial color scheme representing intensity was used for visual clarity.}
\label{fig:simExp_allMeasurements}
\end{figure*}

A schematic of our experimental setup is depicted in \figref{fig:expSetup}. We generate entangled photon pairs using type-II spontaneous parametric down-conversion in a Sagnac interferometer \cite{PhysRevA.101.043815}. We pump a $10$~mm long periodically-poled potassium titanyl phosphate crystal (ppKTP) with a $404$~nm continuous wave diode laser to produce signal and idler photon pairs, both centered at $808$~nm with a spectral bandwidth (FWHM) of approximately $0.4$~nm. The outputs of the interferometer are coupled into single-mode fibers. Immediately following the polarization-entangled source, we measured a $\vert \Phi^+ \rangle$ polarization state fidelity of $96\%$. The signal photons are first sent through an optical telescope to be magnified by a factor of $8.3$, followed by two sets of LOV prism pairs. The magnification controls the number of lattice periods in the emerging intensity pattern by illuminating a larger portion of the prisms. 

The modified signal photons are sent through polarization analyzing optics which consist of a half-wave plate (HWP), a quarter-wave plate (QWP) and a polarizing beam splitter (PBS). Finally, we demagnify the beam by a factor of $4$ by means of a second optical telescope and send the signal photons to an emICCD camera (PI-Max4: 1024 EMB). The idler photons are directly sent to polarization analyzing optics and detected by an avalanche photodiode which triggers the emICCD. Signal photons pass through a $30$~m spool of single-mode fiber in order to compensate for electronic delay. Once the idler photon triggers the camera, an electronic gate in the emICCD collects data for $3$~ns. We measure all 16 combinations of the tomographically complete set $\vert H \rangle$, $\vert V \rangle$, $\vert D \rangle$, and $\vert R \rangle$ on the signal and idler photons. For each polarization measurement, we accumulate signal photons for $2000$ exposures and trigger the camera at a rate of $15$~kHz. Every exposure takes about $2.35$~sec to record. We focus on a $140\times140$~pixel area on the camera, where each pixel is $13$~$\mu$\nobreak m $\times 13$~$\mu$\nobreak m. 


\section{Experimental Results}

\begin{figure*}[ht!]
\centering
\includegraphics[trim={6cm 2.5cm 2.5cm 1.3cm},width=\textwidth]{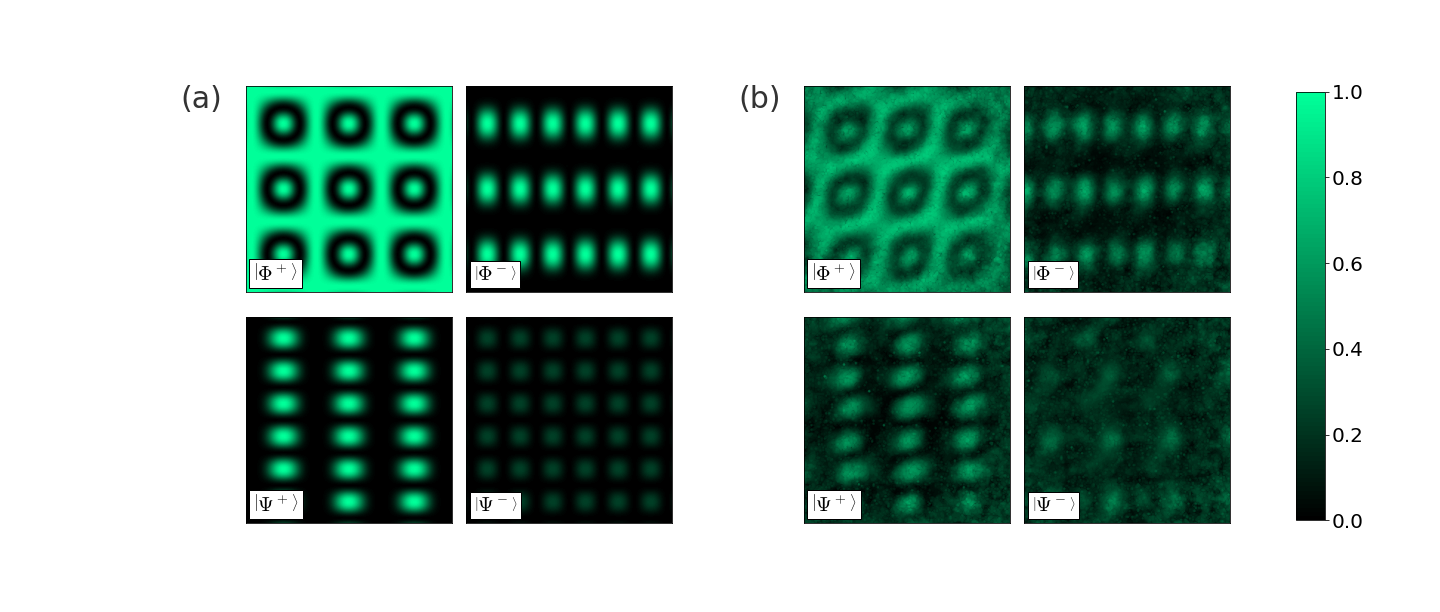}
\caption{Plots of pixel-wise maximum likelihood tomography by means of the fidelity of all four Bell states. (a) Tomography seeded with simulated intensity distributions shown in \figref{fig:simExp_allMeasurements}(a). (b) Tomography seeded with experimental intensity distributions shown in \figref{fig:simExp_allMeasurements}(b). In both cases, $\vert \Phi^+ \rangle$, $\vert \Phi^- \rangle$, $\vert \Psi^+ \rangle$, and $\vert \Psi^- \rangle$ Bell state fidelities are shown. There is good qualitative agreement between experiment and theory, with a reduced experimental fidelity overall.}
\label{fig:Bellstatefidelities}
\end{figure*}

In Fig.~\ref{fig:simExp_allMeasurements}, we show a comparison of theoretically calculated and experimentally measured two-dimensional intensity patterns for all 16 measurement configurations. The theoretical predictions in \figref{fig:simExp_allMeasurements}(a) and the experimental data in \figref{fig:simExp_allMeasurements}(b) are in qualitative agreement. LOV prism pair alignment challenges associated with setting and maintaining the phase leads to slight pattern distortion as compared with theory. In both cases, we used a grid of $140 \times 140$ points. In the image plane of the emICCD, the simulated lattice spacing in \figref{fig:simExp_allMeasurements}(a) is $0.519 \pm 0.015$~mm, while the measured lattice spacing in \figref{fig:simExp_allMeasurements}(b) is $0.522 \pm 0.013$~mm. For the purpose of computing the density matrices, the raw counts from the sum of exposures are used. However, when viewing the intensity distributions, the raw intensity profiles are post-processed using background subtraction and an adaptive two-dimensional Gaussian image filter.

We take the theoretical (\figref{fig:Bellstatefidelities}(a)) and experimental (\figref{fig:Bellstatefidelities}(b)) density matrices  calculated at each pixel position and present the fidelity with each of the four Bell states. For example, the top left image in \figref{fig:Bellstatefidelities}(a) shows how similar the theoretical density matrices, $\rho(x,y)$, are to the $\vert \Phi^+ \rangle$ Bell state by plotting the fidelity, $F(x,y) = \text{Tr}(\rho(x,y)\vert \Phi^+ \rangle \langle \Phi^+ \vert)$, at every pixel position $(x,y)$. The input to the LOV prism pairs is the $\vert \Phi^+ \rangle$ Bell state as shown in \eqref{eq:LOVState}, and you can see from the top left images in \figref{fig:Bellstatefidelities}(a) and \figref{fig:Bellstatefidelities}(b) that the areas around the ring-shaped regions, along with the centre of these regions, have had a phase rotation of a multiple of $2\pi$ from the starting $\vert \Phi^+ \rangle$ Bell state. Looking at the four quadrants of \figref{fig:Bellstatefidelities}(a) and \figref{fig:Bellstatefidelities}(b), it is apparent that at different pixel positions, the input state has been rotated to other Bell states. Pixel-wise quantum state tomography therefore enables a visualization technique to show how the spin-orbit lattice state evolves across the transverse beam profile. The theoretical and experimental Bell state fidelities plotted in \figref{fig:Bellstatefidelities} are in qualitative agreement, and there is a reduced experimental fidelity across all pixels. 

\begin{figure}[ht!]
\centering
\includegraphics[trim={0 0 0 1cm},scale=0.1,width=0.9\columnwidth]{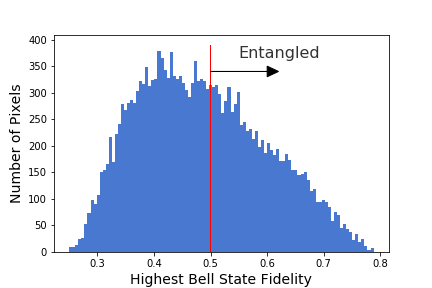}
\caption{Histogram of the highest Bell state fidelity over all pixel positions. A red line is overlaid at $0.5$ fidelity. All pixels with a fidelity greater than $0.5$ with one of the four Bell states are definitely entangled. Using this metric, $42.5\%$ of all pixels in the camera's region of interest are entangled. }
\label{fig:histogram}
\end{figure}

A histogram presenting the highest Bell state fidelity at each pixel position is presented in \figref{fig:histogram}. In the experimental case, 42.5\% of pixel locations have a fidelity of more than 0.5 with one of the four Bell states. This is a witness of entanglement between the signal photons measured at the pixel locations and the idler photons that trigger the camera because qubit separable states cannot achieve a Bell state fidelity of more than 0.5~\cite{RevModPhys.81.865}. In the theoretical case, 85.7\% of pixel locations are a witness of entanglement in this way, so even with perfect image contrast and quantum state preparation, not all positions of this pattern significantly overlap with one of the four Bell states. However, plotting Bell state fidelities helps to illustrate the spatially-dependent rotation of the two-photon spin-orbit lattice state. 

\section{Conclusion}

In this work, we report on the implementation of a remotely prepared optical lattice of spin-orbit states by means of polarization-entangled photon pairs. We experimentally verify the successful remote preparation of this spin-orbit entangled state with an emICCD camera using a pixel-wise maximum likelihood quantum state tomography algorithm. We observe that the entanglement present in the joint two-photon quantum state transforms such that there are overlaps with different Bell states depending on which portion of the LOV prism pairs the signal photon travels through. Furthermore, we have shown that pixel-wise tomography on images acquired by an emICCD camera provides a useful method for observing spatially-dependent two-photon states.

In future work, we plan to study lattices with a higher number of LOV prism pairs to access higher radial quantum numbers and thus a larger alphabet to encode spin-orbit states for quantum communication protocols. Our work advances the study of quantum correlations of structured beams with lattice frameworks, as well as quantum sensing and control of periodic structures where we can take advantage of the novel lattice patterns of our spin-orbit states.

\section*{Acknowledgments}
The authors would like to thank Katanya Kuntz, Ruoxuan Xu, and Jean-Philippe MacLean for helpful discussions and tips. This research was supported in part by the Canadian Excellence Research Chairs (CERC) program, the Natural Sciences and Engineering Research Council of Canada (NSERC), Canada Research Chairs, Industry Canada and the Canada Foundation for Innovation (CFI), Ontario Research Fund (ORF), and the Canada First Research Excellence Fund (CFREF).

\bibliography{OAM_lattices_RSP_ARC}

\begin{thebibliography}{45}%
\makeatletter
\providecommand \@ifxundefined [1]{%
 \@ifx{#1\undefined}
}%
\providecommand \@ifnum [1]{%
 \ifnum #1\expandafter \@firstoftwo
 \else \expandafter \@secondoftwo
 \fi
}%
\providecommand \@ifx [1]{%
 \ifx #1\expandafter \@firstoftwo
 \else \expandafter \@secondoftwo
 \fi
}%
\providecommand \natexlab [1]{#1}%
\providecommand \enquote  [1]{``#1''}%
\providecommand \bibnamefont  [1]{#1}%
\providecommand \bibfnamefont [1]{#1}%
\providecommand \citenamefont [1]{#1}%
\providecommand \href@noop [0]{\@secondoftwo}%
\providecommand \href [0]{\begingroup \@sanitize@url \@href}%
\providecommand \@href[1]{\@@startlink{#1}\@@href}%
\providecommand \@@href[1]{\endgroup#1\@@endlink}%
\providecommand \@sanitize@url [0]{\catcode `\\12\catcode `\$12\catcode
  `\&12\catcode `\#12\catcode `\^12\catcode `\_12\catcode `\%12\relax}%
\providecommand \@@startlink[1]{}%
\providecommand \@@endlink[0]{}%
\providecommand \url  [0]{\begingroup\@sanitize@url \@url }%
\providecommand \@url [1]{\endgroup\@href {#1}{\urlprefix }}%
\providecommand \urlprefix  [0]{URL }%
\providecommand \Eprint [0]{\href }%
\providecommand \doibase [0]{http://dx.doi.org/}%
\providecommand \selectlanguage [0]{\@gobble}%
\providecommand \bibinfo  [0]{\@secondoftwo}%
\providecommand \bibfield  [0]{\@secondoftwo}%
\providecommand \translation [1]{[#1]}%
\providecommand \BibitemOpen [0]{}%
\providecommand \bibitemStop [0]{}%
\providecommand \bibitemNoStop [0]{.\EOS\space}%
\providecommand \EOS [0]{\spacefactor3000\relax}%
\providecommand \BibitemShut  [1]{\csname bibitem#1\endcsname}%
\let\auto@bib@innerbib\@empty
\bibitem [{\citenamefont {Sarenac}\ \emph {et~al.}(2019)\citenamefont
  {Sarenac}, \citenamefont {Kapahi}, \citenamefont {Chen}, \citenamefont
  {Clark}, \citenamefont {Cory}, \citenamefont {Huber}, \citenamefont
  {Taminiau}, \citenamefont {Zhernenkov},\ and\ \citenamefont
  {Pushin}}]{Sarenac20328}%
  \BibitemOpen
  \bibfield  {author} {\bibinfo {author} {\bibfnamefont {D.}~\bibnamefont
  {Sarenac}}, \bibinfo {author} {\bibfnamefont {C.}~\bibnamefont {Kapahi}},
  \bibinfo {author} {\bibfnamefont {W.}~\bibnamefont {Chen}}, \bibinfo {author}
  {\bibfnamefont {C.~W.}\ \bibnamefont {Clark}}, \bibinfo {author}
  {\bibfnamefont {D.~G.}\ \bibnamefont {Cory}}, \bibinfo {author}
  {\bibfnamefont {M.~G.}\ \bibnamefont {Huber}}, \bibinfo {author}
  {\bibfnamefont {I.}~\bibnamefont {Taminiau}}, \bibinfo {author}
  {\bibfnamefont {K.}~\bibnamefont {Zhernenkov}}, \ and\ \bibinfo {author}
  {\bibfnamefont {D.~A.}\ \bibnamefont {Pushin}},\ }\href {\doibase
  10.1073/pnas.1906861116} {\bibfield  {journal} {\bibinfo  {journal}
  {Proceedings of the National Academy of Sciences}\ }\textbf {\bibinfo
  {volume} {116}},\ \bibinfo {pages} {20328} (\bibinfo {year}
  {2019})}\BibitemShut {NoStop}%
\bibitem [{\citenamefont {Karimi}\ \emph {et~al.}(2012)\citenamefont {Karimi},
  \citenamefont {Marrucci}, \citenamefont {Grillo},\ and\ \citenamefont
  {Santamato}}]{matter1}%
  \BibitemOpen
  \bibfield  {author} {\bibinfo {author} {\bibfnamefont {E.}~\bibnamefont
  {Karimi}}, \bibinfo {author} {\bibfnamefont {L.}~\bibnamefont {Marrucci}},
  \bibinfo {author} {\bibfnamefont {V.}~\bibnamefont {Grillo}}, \ and\ \bibinfo
  {author} {\bibfnamefont {E.}~\bibnamefont {Santamato}},\ }\href {\doibase
  10.1103/PhysRevLett.108.044801} {\bibfield  {journal} {\bibinfo  {journal}
  {Phys. Rev. Lett.}\ }\textbf {\bibinfo {volume} {108}},\ \bibinfo {pages}
  {044801} (\bibinfo {year} {2012})}\BibitemShut {NoStop}%
\bibitem [{\citenamefont {Sarenac}\ \emph
  {et~al.}(2018{\natexlab{a}})\citenamefont {Sarenac}, \citenamefont {Nsofini},
  \citenamefont {Hincks}, \citenamefont {Arif}, \citenamefont {Clark},
  \citenamefont {Cory}, \citenamefont {Huber},\ and\ \citenamefont
  {Pushin}}]{sarenac2018methods}%
  \BibitemOpen
  \bibfield  {author} {\bibinfo {author} {\bibfnamefont {D.}~\bibnamefont
  {Sarenac}}, \bibinfo {author} {\bibfnamefont {J.}~\bibnamefont {Nsofini}},
  \bibinfo {author} {\bibfnamefont {I.}~\bibnamefont {Hincks}}, \bibinfo
  {author} {\bibfnamefont {M.}~\bibnamefont {Arif}}, \bibinfo {author}
  {\bibfnamefont {C.~W.}\ \bibnamefont {Clark}}, \bibinfo {author}
  {\bibfnamefont {D.~G.}\ \bibnamefont {Cory}}, \bibinfo {author}
  {\bibfnamefont {M.~G.}\ \bibnamefont {Huber}}, \ and\ \bibinfo {author}
  {\bibfnamefont {D.~A.}\ \bibnamefont {Pushin}},\ }\href {\doibase
  10.1088/1367-2630/aae3ac} {\bibfield  {journal} {\bibinfo  {journal} {New
  Journal of Physics}\ }\textbf {\bibinfo {volume} {20}},\ \bibinfo {pages}
  {103012} (\bibinfo {year} {2018}{\natexlab{a}})}\BibitemShut {NoStop}%
\bibitem [{\citenamefont {Rubinsztein-Dunlop}\ \emph
  {et~al.}(2016)\citenamefont {Rubinsztein-Dunlop}, \citenamefont {Forbes},
  \citenamefont {Berry}, \citenamefont {Dennis}, \citenamefont {Andrews},
  \citenamefont {Mansuripur}, \citenamefont {Denz}, \citenamefont {Alpmann},
  \citenamefont {Banzer}, \citenamefont {Bauer}, \citenamefont {Karimi},
  \citenamefont {Marrucci}, \citenamefont {Padgett}, \citenamefont
  {Ritsch-Marte}, \citenamefont {Litchinitser}, \citenamefont {Bigelow},
  \citenamefont {Rosales-Guzm{\'{a}}n}, \citenamefont {Belmonte}, \citenamefont
  {Torres}, \citenamefont {Neely}, \citenamefont {Baker}, \citenamefont
  {Gordon}, \citenamefont {Stilgoe}, \citenamefont {Romero}, \citenamefont
  {White}, \citenamefont {Fickler}, \citenamefont {Willner}, \citenamefont
  {Xie}, \citenamefont {McMorran},\ and\ \citenamefont
  {Weiner}}]{Rubinsztein_Dunlop_2016}%
  \BibitemOpen
  \bibfield  {author} {\bibinfo {author} {\bibfnamefont {H.}~\bibnamefont
  {Rubinsztein-Dunlop}}, \bibinfo {author} {\bibfnamefont {A.}~\bibnamefont
  {Forbes}}, \bibinfo {author} {\bibfnamefont {M.~V.}\ \bibnamefont {Berry}},
  \bibinfo {author} {\bibfnamefont {M.~R.}\ \bibnamefont {Dennis}}, \bibinfo
  {author} {\bibfnamefont {D.~L.}\ \bibnamefont {Andrews}}, \bibinfo {author}
  {\bibfnamefont {M.}~\bibnamefont {Mansuripur}}, \bibinfo {author}
  {\bibfnamefont {C.}~\bibnamefont {Denz}}, \bibinfo {author} {\bibfnamefont
  {C.}~\bibnamefont {Alpmann}}, \bibinfo {author} {\bibfnamefont
  {P.}~\bibnamefont {Banzer}}, \bibinfo {author} {\bibfnamefont
  {T.}~\bibnamefont {Bauer}}, \bibinfo {author} {\bibfnamefont
  {E.}~\bibnamefont {Karimi}}, \bibinfo {author} {\bibfnamefont
  {L.}~\bibnamefont {Marrucci}}, \bibinfo {author} {\bibfnamefont
  {M.}~\bibnamefont {Padgett}}, \bibinfo {author} {\bibfnamefont
  {M.}~\bibnamefont {Ritsch-Marte}}, \bibinfo {author} {\bibfnamefont {N.~M.}\
  \bibnamefont {Litchinitser}}, \bibinfo {author} {\bibfnamefont {N.~P.}\
  \bibnamefont {Bigelow}}, \bibinfo {author} {\bibfnamefont {C.}~\bibnamefont
  {Rosales-Guzm{\'{a}}n}}, \bibinfo {author} {\bibfnamefont {A.}~\bibnamefont
  {Belmonte}}, \bibinfo {author} {\bibfnamefont {J.~P.}\ \bibnamefont
  {Torres}}, \bibinfo {author} {\bibfnamefont {T.~W.}\ \bibnamefont {Neely}},
  \bibinfo {author} {\bibfnamefont {M.}~\bibnamefont {Baker}}, \bibinfo
  {author} {\bibfnamefont {R.}~\bibnamefont {Gordon}}, \bibinfo {author}
  {\bibfnamefont {A.~B.}\ \bibnamefont {Stilgoe}}, \bibinfo {author}
  {\bibfnamefont {J.}~\bibnamefont {Romero}}, \bibinfo {author} {\bibfnamefont
  {A.~G.}\ \bibnamefont {White}}, \bibinfo {author} {\bibfnamefont
  {R.}~\bibnamefont {Fickler}}, \bibinfo {author} {\bibfnamefont {A.~E.}\
  \bibnamefont {Willner}}, \bibinfo {author} {\bibfnamefont {G.}~\bibnamefont
  {Xie}}, \bibinfo {author} {\bibfnamefont {B.}~\bibnamefont {McMorran}}, \
  and\ \bibinfo {author} {\bibfnamefont {A.~M.}\ \bibnamefont {Weiner}},\
  }\href {\doibase 10.1088/2040-8978/19/1/013001} {\bibfield  {journal}
  {\bibinfo  {journal} {Journal of Optics}\ }\textbf {\bibinfo {volume} {19}},\
  \bibinfo {pages} {013001} (\bibinfo {year} {2016})}\BibitemShut {NoStop}%
\bibitem [{\citenamefont {Uchida}\ and\ \citenamefont
  {Tonomura}(2010)}]{matter2}%
  \BibitemOpen
  \bibfield  {author} {\bibinfo {author} {\bibfnamefont {M.}~\bibnamefont
  {Uchida}}\ and\ \bibinfo {author} {\bibfnamefont {A.}~\bibnamefont
  {Tonomura}},\ }\href {\doibase 10.1038/nature08904} {\bibfield  {journal}
  {\bibinfo  {journal} {Nature}\ }\textbf {\bibinfo {volume} {464}},\ \bibinfo
  {pages} {737} (\bibinfo {year} {2010})}\BibitemShut {NoStop}%
\bibitem [{\citenamefont {Ikonnikov}\ \emph {et~al.}(2020)\citenamefont
  {Ikonnikov}, \citenamefont {Myslivets}, \citenamefont {Volochaev},
  \citenamefont {Arkhipkin},\ and\ \citenamefont {Vyunishev}}]{talbot2}%
  \BibitemOpen
  \bibfield  {author} {\bibinfo {author} {\bibfnamefont {D.~A.}\ \bibnamefont
  {Ikonnikov}}, \bibinfo {author} {\bibfnamefont {S.~A.}\ \bibnamefont
  {Myslivets}}, \bibinfo {author} {\bibfnamefont {M.~N.}\ \bibnamefont
  {Volochaev}}, \bibinfo {author} {\bibfnamefont {V.~G.}\ \bibnamefont
  {Arkhipkin}}, \ and\ \bibinfo {author} {\bibfnamefont {A.~M.}\ \bibnamefont
  {Vyunishev}},\ }\href {\doibase 10.1038/s41598-020-77418-y} {\bibfield
  {journal} {\bibinfo  {journal} {Scientific Reports}\ }\textbf {\bibinfo
  {volume} {10}},\ \bibinfo {pages} {20315} (\bibinfo {year}
  {2020})}\BibitemShut {NoStop}%
\bibitem [{\citenamefont {Schwarz}\ \emph {et~al.}(2020)\citenamefont
  {Schwarz}, \citenamefont {Kapahi}, \citenamefont {Xu}, \citenamefont
  {Cameron}, \citenamefont {Sarenac}, \citenamefont {MacLean}, \citenamefont
  {Kuntz}, \citenamefont {Cory}, \citenamefont {Jennewein}, \citenamefont
  {Resch},\ and\ \citenamefont {Pushin}}]{PhysRevA.101.043815}%
  \BibitemOpen
  \bibfield  {author} {\bibinfo {author} {\bibfnamefont {S.}~\bibnamefont
  {Schwarz}}, \bibinfo {author} {\bibfnamefont {C.}~\bibnamefont {Kapahi}},
  \bibinfo {author} {\bibfnamefont {R.}~\bibnamefont {Xu}}, \bibinfo {author}
  {\bibfnamefont {A.~R.}\ \bibnamefont {Cameron}}, \bibinfo {author}
  {\bibfnamefont {D.}~\bibnamefont {Sarenac}}, \bibinfo {author} {\bibfnamefont
  {J.~P.~W.}\ \bibnamefont {MacLean}}, \bibinfo {author} {\bibfnamefont
  {K.~B.}\ \bibnamefont {Kuntz}}, \bibinfo {author} {\bibfnamefont {D.~G.}\
  \bibnamefont {Cory}}, \bibinfo {author} {\bibfnamefont {T.}~\bibnamefont
  {Jennewein}}, \bibinfo {author} {\bibfnamefont {K.~J.}\ \bibnamefont
  {Resch}}, \ and\ \bibinfo {author} {\bibfnamefont {D.~A.}\ \bibnamefont
  {Pushin}},\ }\href {\doibase 10.1103/PhysRevA.101.043815} {\bibfield
  {journal} {\bibinfo  {journal} {Phys. Rev. A}\ }\textbf {\bibinfo {volume}
  {101}},\ \bibinfo {pages} {043815} (\bibinfo {year} {2020})}\BibitemShut
  {NoStop}%
\bibitem [{\citenamefont {Li}\ \emph {et~al.}(2018)\citenamefont {Li},
  \citenamefont {Ma}, \citenamefont {Zhang}, \citenamefont {Tai}, \citenamefont
  {Li}, \citenamefont {Tang}, \citenamefont {Wang}, \citenamefont {Tang},\ and\
  \citenamefont {Cai}}]{optical_lattice_shaping1}%
  \BibitemOpen
  \bibfield  {author} {\bibinfo {author} {\bibfnamefont {X.}~\bibnamefont
  {Li}}, \bibinfo {author} {\bibfnamefont {H.}~\bibnamefont {Ma}}, \bibinfo
  {author} {\bibfnamefont {H.}~\bibnamefont {Zhang}}, \bibinfo {author}
  {\bibfnamefont {Y.}~\bibnamefont {Tai}}, \bibinfo {author} {\bibfnamefont
  {H.}~\bibnamefont {Li}}, \bibinfo {author} {\bibfnamefont {M.}~\bibnamefont
  {Tang}}, \bibinfo {author} {\bibfnamefont {J.}~\bibnamefont {Wang}}, \bibinfo
  {author} {\bibfnamefont {J.}~\bibnamefont {Tang}}, \ and\ \bibinfo {author}
  {\bibfnamefont {Y.}~\bibnamefont {Cai}},\ }\href {\doibase
  10.1364/OE.26.022965} {\bibfield  {journal} {\bibinfo  {journal} {Opt.
  Express}\ }\textbf {\bibinfo {volume} {26}},\ \bibinfo {pages} {22965}
  (\bibinfo {year} {2018})}\BibitemShut {NoStop}%
\bibitem [{\citenamefont {Han}\ \emph {et~al.}(2019)\citenamefont {Han},
  \citenamefont {Rong}, \citenamefont {Zhang},\ and\ \citenamefont
  {Chen}}]{optical_lattice_shaping2}%
  \BibitemOpen
  \bibfield  {author} {\bibinfo {author} {\bibfnamefont {Y.-J.}\ \bibnamefont
  {Han}}, \bibinfo {author} {\bibfnamefont {Z.-Y.}\ \bibnamefont {Rong}},
  \bibinfo {author} {\bibfnamefont {L.}~\bibnamefont {Zhang}}, \ and\ \bibinfo
  {author} {\bibfnamefont {X.-Y.}\ \bibnamefont {Chen}},\ }\href {\doibase
  10.1364/AO.58.006325} {\bibfield  {journal} {\bibinfo  {journal} {Appl.
  Opt.}\ }\textbf {\bibinfo {volume} {58}},\ \bibinfo {pages} {6325} (\bibinfo
  {year} {2019})}\BibitemShut {NoStop}%
\bibitem [{\citenamefont {Sarenac}\ \emph
  {et~al.}(2020{\natexlab{a}})\citenamefont {Sarenac}, \citenamefont {Kapahi},
  \citenamefont {Silva}, \citenamefont {Cory}, \citenamefont {Taminiau},
  \citenamefont {Thompson},\ and\ \citenamefont {Pushin}}]{Sarenac14682}%
  \BibitemOpen
  \bibfield  {author} {\bibinfo {author} {\bibfnamefont {D.}~\bibnamefont
  {Sarenac}}, \bibinfo {author} {\bibfnamefont {C.}~\bibnamefont {Kapahi}},
  \bibinfo {author} {\bibfnamefont {A.~E.}\ \bibnamefont {Silva}}, \bibinfo
  {author} {\bibfnamefont {D.~G.}\ \bibnamefont {Cory}}, \bibinfo {author}
  {\bibfnamefont {I.}~\bibnamefont {Taminiau}}, \bibinfo {author}
  {\bibfnamefont {B.}~\bibnamefont {Thompson}}, \ and\ \bibinfo {author}
  {\bibfnamefont {D.~A.}\ \bibnamefont {Pushin}},\ }\href {\doibase
  10.1073/pnas.1920226117} {\bibfield  {journal} {\bibinfo  {journal}
  {Proceedings of the National Academy of Sciences}\ }\textbf {\bibinfo
  {volume} {117}},\ \bibinfo {pages} {14682} (\bibinfo {year}
  {2020}{\natexlab{a}})}\BibitemShut {NoStop}%
\bibitem [{\citenamefont {Sarenac}\ \emph
  {et~al.}(2020{\natexlab{b}})\citenamefont {Sarenac}, \citenamefont {Silva},
  \citenamefont {Kapahi}, \citenamefont {Thompson}, \citenamefont {Cory},\ and\
  \citenamefont {Pushin}}]{sarenac2020human}%
  \BibitemOpen
  \bibfield  {author} {\bibinfo {author} {\bibfnamefont {D.}~\bibnamefont
  {Sarenac}}, \bibinfo {author} {\bibfnamefont {A.~E.}\ \bibnamefont {Silva}},
  \bibinfo {author} {\bibfnamefont {C.}~\bibnamefont {Kapahi}}, \bibinfo
  {author} {\bibfnamefont {B.}~\bibnamefont {Thompson}}, \bibinfo {author}
  {\bibfnamefont {D.~G.}\ \bibnamefont {Cory}}, \ and\ \bibinfo {author}
  {\bibfnamefont {D.~A.}\ \bibnamefont {Pushin}},\ }\href@noop {} {\enquote
  {\bibinfo {title} {Human psychophysical discrimination of spatially dependant
  pancharatnam-berry phases in optical spin-orbit states},}\ } (\bibinfo {year}
  {2020}{\natexlab{b}}),\ \Eprint {http://arxiv.org/abs/2010.09619}
  {arXiv:2010.09619 [physics.optics]} \BibitemShut {NoStop}%
\bibitem [{\citenamefont {Andersen}\ \emph {et~al.}(2006)\citenamefont
  {Andersen}, \citenamefont {Ryu}, \citenamefont {Clade}, \citenamefont
  {Natarajan}, \citenamefont {Vaziri}, \citenamefont {Helmerson},\ and\
  \citenamefont {Phillips}}]{Andersen2006}%
  \BibitemOpen
  \bibfield  {author} {\bibinfo {author} {\bibfnamefont {M.~F.}\ \bibnamefont
  {Andersen}}, \bibinfo {author} {\bibfnamefont {C.}~\bibnamefont {Ryu}},
  \bibinfo {author} {\bibfnamefont {P.}~\bibnamefont {Clade}}, \bibinfo
  {author} {\bibfnamefont {V.}~\bibnamefont {Natarajan}}, \bibinfo {author}
  {\bibfnamefont {A.}~\bibnamefont {Vaziri}}, \bibinfo {author} {\bibfnamefont
  {K.}~\bibnamefont {Helmerson}}, \ and\ \bibinfo {author} {\bibfnamefont
  {W.~D.}\ \bibnamefont {Phillips}},\ }\href {\doibase
  10.1103/PhysRevLett.97.170406} {\bibfield  {journal} {\bibinfo  {journal}
  {Phys. Rev. Lett.}\ }\textbf {\bibinfo {volume} {97}},\ \bibinfo {pages}
  {170406} (\bibinfo {year} {2006})}\BibitemShut {NoStop}%
\bibitem [{\citenamefont {He}\ \emph {et~al.}(1995)\citenamefont {He},
  \citenamefont {Friese}, \citenamefont {Heckenberg},\ and\ \citenamefont
  {Rubinsztein-Dunlop}}]{he1995direct}%
  \BibitemOpen
  \bibfield  {author} {\bibinfo {author} {\bibfnamefont {H.}~\bibnamefont
  {He}}, \bibinfo {author} {\bibfnamefont {M.~E.~J.}\ \bibnamefont {Friese}},
  \bibinfo {author} {\bibfnamefont {N.~R.}\ \bibnamefont {Heckenberg}}, \ and\
  \bibinfo {author} {\bibfnamefont {H.}~\bibnamefont {Rubinsztein-Dunlop}},\
  }\href {\doibase 10.1103/PhysRevLett.75.826} {\bibfield  {journal} {\bibinfo
  {journal} {Phys. Rev. Lett.}\ }\textbf {\bibinfo {volume} {75}},\ \bibinfo
  {pages} {826} (\bibinfo {year} {1995})}\BibitemShut {NoStop}%
\bibitem [{\citenamefont {Schmiegelow}\ and\ \citenamefont
  {Schmidt-Kaler}(2012)}]{Schmiegelow2012}%
  \BibitemOpen
  \bibfield  {author} {\bibinfo {author} {\bibfnamefont {C.}~\bibnamefont
  {Schmiegelow}}\ and\ \bibinfo {author} {\bibfnamefont {F.}~\bibnamefont
  {Schmidt-Kaler}},\ }\href {\doibase 10.1140/epjd/e2012-20730-4} {\bibfield
  {journal} {\bibinfo  {journal} {The European Physical Journal D}\ }\textbf
  {\bibinfo {volume} {66}},\ \bibinfo {pages} {157} (\bibinfo {year}
  {2012})}\BibitemShut {NoStop}%
\bibitem [{\citenamefont {Luo}\ \emph {et~al.}(2017)\citenamefont {Luo},
  \citenamefont {Zhou}, \citenamefont {Xu}, \citenamefont {Li}, \citenamefont
  {Guo}, \citenamefont {Zhang},\ and\ \citenamefont
  {Zhou}}]{luo_synthetic-lattice_2017}%
  \BibitemOpen
  \bibfield  {author} {\bibinfo {author} {\bibfnamefont {X.-W.}\ \bibnamefont
  {Luo}}, \bibinfo {author} {\bibfnamefont {X.}~\bibnamefont {Zhou}}, \bibinfo
  {author} {\bibfnamefont {J.-S.}\ \bibnamefont {Xu}}, \bibinfo {author}
  {\bibfnamefont {C.-F.}\ \bibnamefont {Li}}, \bibinfo {author} {\bibfnamefont
  {G.-C.}\ \bibnamefont {Guo}}, \bibinfo {author} {\bibfnamefont
  {C.}~\bibnamefont {Zhang}}, \ and\ \bibinfo {author} {\bibfnamefont {Z.-W.}\
  \bibnamefont {Zhou}},\ }\href {\doibase 10.1038/ncomms16097} {\bibfield
  {journal} {\bibinfo  {journal} {Nature Communications}\ }\textbf {\bibinfo
  {volume} {8}},\ \bibinfo {pages} {16097} (\bibinfo {year}
  {2017})}\BibitemShut {NoStop}%
\bibitem [{\citenamefont {Morrow}\ \emph {et~al.}(2002)\citenamefont {Morrow},
  \citenamefont {Dutta},\ and\ \citenamefont {Raithel}}]{optical_lattice}%
  \BibitemOpen
  \bibfield  {author} {\bibinfo {author} {\bibfnamefont {N.~V.}\ \bibnamefont
  {Morrow}}, \bibinfo {author} {\bibfnamefont {S.~K.}\ \bibnamefont {Dutta}}, \
  and\ \bibinfo {author} {\bibfnamefont {G.}~\bibnamefont {Raithel}},\ }\href
  {\doibase 10.1103/PhysRevLett.88.093003} {\bibfield  {journal} {\bibinfo
  {journal} {Phys. Rev. Lett.}\ }\textbf {\bibinfo {volume} {88}},\ \bibinfo
  {pages} {093003} (\bibinfo {year} {2002})}\BibitemShut {NoStop}%
\bibitem [{\citenamefont {Forbes}\ and\ \citenamefont
  {Nape}(2019)}]{Forbes2019}%
  \BibitemOpen
  \bibfield  {author} {\bibinfo {author} {\bibfnamefont {A.}~\bibnamefont
  {Forbes}}\ and\ \bibinfo {author} {\bibfnamefont {I.}~\bibnamefont {Nape}},\
  }\href {\doibase 10.1116/1.5112027} {\bibfield  {journal} {\bibinfo
  {journal} {AVS Quantum Science}\ }\textbf {\bibinfo {volume} {1}},\ \bibinfo
  {pages} {011701} (\bibinfo {year} {2019})}\BibitemShut {NoStop}%
\bibitem [{\citenamefont {Barreiro}\ \emph {et~al.}(2008)\citenamefont
  {Barreiro}, \citenamefont {Wei},\ and\ \citenamefont {Kwiat}}]{Barreiro2008}%
  \BibitemOpen
  \bibfield  {author} {\bibinfo {author} {\bibfnamefont {J.~T.}\ \bibnamefont
  {Barreiro}}, \bibinfo {author} {\bibfnamefont {T.-C.}\ \bibnamefont {Wei}}, \
  and\ \bibinfo {author} {\bibfnamefont {P.~G.}\ \bibnamefont {Kwiat}},\ }\href
  {\doibase 10.1038/nphys919} {\bibfield  {journal} {\bibinfo  {journal}
  {Nature Physics}\ }\textbf {\bibinfo {volume} {4}},\ \bibinfo {pages} {282}
  (\bibinfo {year} {2008})}\BibitemShut {NoStop}%
\bibitem [{\citenamefont {Marrucci}\ \emph {et~al.}(2011)\citenamefont
  {Marrucci}, \citenamefont {Karimi}, \citenamefont {Slussarenko},
  \citenamefont {Piccirillo}, \citenamefont {Santamato}, \citenamefont
  {Nagali},\ and\ \citenamefont {Sciarrino}}]{Marrucci2011}%
  \BibitemOpen
  \bibfield  {author} {\bibinfo {author} {\bibfnamefont {L.}~\bibnamefont
  {Marrucci}}, \bibinfo {author} {\bibfnamefont {E.}~\bibnamefont {Karimi}},
  \bibinfo {author} {\bibfnamefont {S.}~\bibnamefont {Slussarenko}}, \bibinfo
  {author} {\bibfnamefont {B.}~\bibnamefont {Piccirillo}}, \bibinfo {author}
  {\bibfnamefont {E.}~\bibnamefont {Santamato}}, \bibinfo {author}
  {\bibfnamefont {E.}~\bibnamefont {Nagali}}, \ and\ \bibinfo {author}
  {\bibfnamefont {F.}~\bibnamefont {Sciarrino}},\ }\href {\doibase
  10.1088/2040-8978/13/6/064001} {\bibfield  {journal} {\bibinfo  {journal}
  {Journal of Optics}\ }\textbf {\bibinfo {volume} {13}},\ \bibinfo {pages}
  {064001} (\bibinfo {year} {2011})}\BibitemShut {NoStop}%
\bibitem [{\citenamefont {Milione}\ \emph {et~al.}(2015)\citenamefont
  {Milione}, \citenamefont {Lavery}, \citenamefont {Huang}, \citenamefont
  {Ren}, \citenamefont {Xie}, \citenamefont {Nguyen}, \citenamefont {Karimi},
  \citenamefont {Marrucci}, \citenamefont {Nolan}, \citenamefont {Alfano},\
  and\ \citenamefont {Willner}}]{Milione2015}%
  \BibitemOpen
  \bibfield  {author} {\bibinfo {author} {\bibfnamefont {G.}~\bibnamefont
  {Milione}}, \bibinfo {author} {\bibfnamefont {M.~P.~J.}\ \bibnamefont
  {Lavery}}, \bibinfo {author} {\bibfnamefont {H.}~\bibnamefont {Huang}},
  \bibinfo {author} {\bibfnamefont {Y.}~\bibnamefont {Ren}}, \bibinfo {author}
  {\bibfnamefont {G.}~\bibnamefont {Xie}}, \bibinfo {author} {\bibfnamefont
  {T.~A.}\ \bibnamefont {Nguyen}}, \bibinfo {author} {\bibfnamefont
  {E.}~\bibnamefont {Karimi}}, \bibinfo {author} {\bibfnamefont
  {L.}~\bibnamefont {Marrucci}}, \bibinfo {author} {\bibfnamefont {D.~A.}\
  \bibnamefont {Nolan}}, \bibinfo {author} {\bibfnamefont {R.~R.}\ \bibnamefont
  {Alfano}}, \ and\ \bibinfo {author} {\bibfnamefont {A.~E.}\ \bibnamefont
  {Willner}},\ }\href {\doibase 10.1364/OL.40.001980} {\bibfield  {journal}
  {\bibinfo  {journal} {Opt. Lett.}\ }\textbf {\bibinfo {volume} {40}},\
  \bibinfo {pages} {1980} (\bibinfo {year} {2015})}\BibitemShut {NoStop}%
\bibitem [{\citenamefont {Vallone}\ \emph {et~al.}(2014)\citenamefont
  {Vallone}, \citenamefont {D'Ambrosio}, \citenamefont {Sponselli},
  \citenamefont {Slussarenko}, \citenamefont {Marrucci}, \citenamefont
  {Sciarrino},\ and\ \citenamefont {Villoresi}}]{Vallone2014}%
  \BibitemOpen
  \bibfield  {author} {\bibinfo {author} {\bibfnamefont {G.}~\bibnamefont
  {Vallone}}, \bibinfo {author} {\bibfnamefont {V.}~\bibnamefont {D'Ambrosio}},
  \bibinfo {author} {\bibfnamefont {A.}~\bibnamefont {Sponselli}}, \bibinfo
  {author} {\bibfnamefont {S.}~\bibnamefont {Slussarenko}}, \bibinfo {author}
  {\bibfnamefont {L.}~\bibnamefont {Marrucci}}, \bibinfo {author}
  {\bibfnamefont {F.}~\bibnamefont {Sciarrino}}, \ and\ \bibinfo {author}
  {\bibfnamefont {P.}~\bibnamefont {Villoresi}},\ }\href {\doibase
  10.1103/PhysRevLett.113.060503} {\bibfield  {journal} {\bibinfo  {journal}
  {Phys. Rev. Lett.}\ }\textbf {\bibinfo {volume} {113}},\ \bibinfo {pages}
  {060503} (\bibinfo {year} {2014})}\BibitemShut {NoStop}%
\bibitem [{\citenamefont {Schmiegelow}\ \emph {et~al.}(2016)\citenamefont
  {Schmiegelow}, \citenamefont {Schulz}, \citenamefont {Kaufmann},
  \citenamefont {Ruster}, \citenamefont {Poschinger},\ and\ \citenamefont
  {Schmidt-Kaler}}]{Schmiegelow2016}%
  \BibitemOpen
  \bibfield  {author} {\bibinfo {author} {\bibfnamefont {C.~T.}\ \bibnamefont
  {Schmiegelow}}, \bibinfo {author} {\bibfnamefont {J.}~\bibnamefont {Schulz}},
  \bibinfo {author} {\bibfnamefont {H.}~\bibnamefont {Kaufmann}}, \bibinfo
  {author} {\bibfnamefont {T.}~\bibnamefont {Ruster}}, \bibinfo {author}
  {\bibfnamefont {U.~G.}\ \bibnamefont {Poschinger}}, \ and\ \bibinfo {author}
  {\bibfnamefont {F.}~\bibnamefont {Schmidt-Kaler}},\ }\href {\doibase
  10.1038/ncomms12998} {\bibfield  {journal} {\bibinfo  {journal} {Nature
  Communications}\ }\textbf {\bibinfo {volume} {7}},\ \bibinfo {pages} {12998}
  (\bibinfo {year} {2016})}\BibitemShut {NoStop}%
\bibitem [{\citenamefont {Erhard}\ \emph {et~al.}(2018)\citenamefont {Erhard},
  \citenamefont {Fickler}, \citenamefont {Krenn},\ and\ \citenamefont
  {Zeilinger}}]{erhard2018twisted}%
  \BibitemOpen
  \bibfield  {author} {\bibinfo {author} {\bibfnamefont {M.}~\bibnamefont
  {Erhard}}, \bibinfo {author} {\bibfnamefont {R.}~\bibnamefont {Fickler}},
  \bibinfo {author} {\bibfnamefont {M.}~\bibnamefont {Krenn}}, \ and\ \bibinfo
  {author} {\bibfnamefont {A.}~\bibnamefont {Zeilinger}},\ }\href
  {https://doi.org/10.1038/lsa.2017.146 http://10.0.4.14/lsa.2017.146}
  {\bibfield  {journal} {\bibinfo  {journal} {Light: Science {\&}Amp;
  Applications}\ }\textbf {\bibinfo {volume} {7}},\ \bibinfo {pages} {17146}
  (\bibinfo {year} {2018})}\BibitemShut {NoStop}%
\bibitem [{\citenamefont {Fickler}\ \emph {et~al.}(2016)\citenamefont
  {Fickler}, \citenamefont {Campbell}, \citenamefont {Buchler}, \citenamefont
  {Lam},\ and\ \citenamefont {Zeilinger}}]{fickler2016}%
  \BibitemOpen
  \bibfield  {author} {\bibinfo {author} {\bibfnamefont {R.}~\bibnamefont
  {Fickler}}, \bibinfo {author} {\bibfnamefont {G.}~\bibnamefont {Campbell}},
  \bibinfo {author} {\bibfnamefont {B.}~\bibnamefont {Buchler}}, \bibinfo
  {author} {\bibfnamefont {P.~K.}\ \bibnamefont {Lam}}, \ and\ \bibinfo
  {author} {\bibfnamefont {A.}~\bibnamefont {Zeilinger}},\ }\href {\doibase
  10.1073/pnas.1616889113} {\bibfield  {journal} {\bibinfo  {journal}
  {Proceedings of the National Academy of Sciences}\ }\textbf {\bibinfo
  {volume} {113}},\ \bibinfo {pages} {13642} (\bibinfo {year}
  {2016})}\BibitemShut {NoStop}%
\bibitem [{\citenamefont {Sit}\ \emph {et~al.}(2017)\citenamefont {Sit},
  \citenamefont {Bouchard}, \citenamefont {Fickler}, \citenamefont
  {Gagnon-Bischoff}, \citenamefont {Larocque}, \citenamefont {Heshami},
  \citenamefont {Elser}, \citenamefont {Peuntinger}, \citenamefont
  {G\"{u}nthner}, \citenamefont {Heim}, \citenamefont {Marquardt},
  \citenamefont {Leuchs}, \citenamefont {Boyd},\ and\ \citenamefont
  {Karimi}}]{Sit2017}%
  \BibitemOpen
  \bibfield  {author} {\bibinfo {author} {\bibfnamefont {A.}~\bibnamefont
  {Sit}}, \bibinfo {author} {\bibfnamefont {F.}~\bibnamefont {Bouchard}},
  \bibinfo {author} {\bibfnamefont {R.}~\bibnamefont {Fickler}}, \bibinfo
  {author} {\bibfnamefont {J.}~\bibnamefont {Gagnon-Bischoff}}, \bibinfo
  {author} {\bibfnamefont {H.}~\bibnamefont {Larocque}}, \bibinfo {author}
  {\bibfnamefont {K.}~\bibnamefont {Heshami}}, \bibinfo {author} {\bibfnamefont
  {D.}~\bibnamefont {Elser}}, \bibinfo {author} {\bibfnamefont
  {C.}~\bibnamefont {Peuntinger}}, \bibinfo {author} {\bibfnamefont
  {K.}~\bibnamefont {G\"{u}nthner}}, \bibinfo {author} {\bibfnamefont
  {B.}~\bibnamefont {Heim}}, \bibinfo {author} {\bibfnamefont {C.}~\bibnamefont
  {Marquardt}}, \bibinfo {author} {\bibfnamefont {G.}~\bibnamefont {Leuchs}},
  \bibinfo {author} {\bibfnamefont {R.~W.}\ \bibnamefont {Boyd}}, \ and\
  \bibinfo {author} {\bibfnamefont {E.}~\bibnamefont {Karimi}},\ }\href
  {\doibase 10.1364/OPTICA.4.001006} {\bibfield  {journal} {\bibinfo  {journal}
  {Optica}\ }\textbf {\bibinfo {volume} {4}},\ \bibinfo {pages} {1006}
  (\bibinfo {year} {2017})}\BibitemShut {NoStop}%
\bibitem [{\citenamefont {Nagali}\ \emph {et~al.}(2009)\citenamefont {Nagali},
  \citenamefont {Sciarrino}, \citenamefont {De~Martini}, \citenamefont
  {Marrucci}, \citenamefont {Piccirillo}, \citenamefont {Karimi},\ and\
  \citenamefont {Santamato}}]{Nagali2009}%
  \BibitemOpen
  \bibfield  {author} {\bibinfo {author} {\bibfnamefont {E.}~\bibnamefont
  {Nagali}}, \bibinfo {author} {\bibfnamefont {F.}~\bibnamefont {Sciarrino}},
  \bibinfo {author} {\bibfnamefont {F.}~\bibnamefont {De~Martini}}, \bibinfo
  {author} {\bibfnamefont {L.}~\bibnamefont {Marrucci}}, \bibinfo {author}
  {\bibfnamefont {B.}~\bibnamefont {Piccirillo}}, \bibinfo {author}
  {\bibfnamefont {E.}~\bibnamefont {Karimi}}, \ and\ \bibinfo {author}
  {\bibfnamefont {E.}~\bibnamefont {Santamato}},\ }\href {\doibase
  10.1103/PhysRevLett.103.013601} {\bibfield  {journal} {\bibinfo  {journal}
  {Phys. Rev. Lett.}\ }\textbf {\bibinfo {volume} {103}},\ \bibinfo {pages}
  {013601} (\bibinfo {year} {2009})}\BibitemShut {NoStop}%
\bibitem [{\citenamefont {Wang}\ \emph {et~al.}(2011)\citenamefont {Wang},
  \citenamefont {Zhang},\ and\ \citenamefont {Zhang}}]{Wang2011}%
  \BibitemOpen
  \bibfield  {author} {\bibinfo {author} {\bibfnamefont {C.}~\bibnamefont
  {Wang}}, \bibinfo {author} {\bibfnamefont {Y.}~\bibnamefont {Zhang}}, \ and\
  \bibinfo {author} {\bibfnamefont {R.}~\bibnamefont {Zhang}},\ }\href
  {\doibase 10.1364/OE.19.025685} {\bibfield  {journal} {\bibinfo  {journal}
  {Opt. Express}\ }\textbf {\bibinfo {volume} {19}},\ \bibinfo {pages} {25685}
  (\bibinfo {year} {2011})}\BibitemShut {NoStop}%
\bibitem [{\citenamefont {Diamanti}\ \emph {et~al.}(2016)\citenamefont
  {Diamanti}, \citenamefont {Lo}, \citenamefont {Qi},\ and\ \citenamefont
  {Yuan}}]{diamanti_practical_2016}%
  \BibitemOpen
  \bibfield  {author} {\bibinfo {author} {\bibfnamefont {E.}~\bibnamefont
  {Diamanti}}, \bibinfo {author} {\bibfnamefont {H.-K.}\ \bibnamefont {Lo}},
  \bibinfo {author} {\bibfnamefont {B.}~\bibnamefont {Qi}}, \ and\ \bibinfo
  {author} {\bibfnamefont {Z.}~\bibnamefont {Yuan}},\ }\href {\doibase
  10.1038/npjqi.2016.25} {\bibfield  {journal} {\bibinfo  {journal} {npj
  Quantum Information}\ }\textbf {\bibinfo {volume} {2}},\ \bibinfo {pages}
  {16025} (\bibinfo {year} {2016})}\BibitemShut {NoStop}%
\bibitem [{\citenamefont {Mafu}\ \emph {et~al.}(2013)\citenamefont {Mafu},
  \citenamefont {Dudley}, \citenamefont {Goyal}, \citenamefont {Giovannini},
  \citenamefont {McLaren}, \citenamefont {Padgett}, \citenamefont {Konrad},
  \citenamefont {Petruccione}, \citenamefont {L\"utkenhaus},\ and\
  \citenamefont {Forbes}}]{Mafu2013}%
  \BibitemOpen
  \bibfield  {author} {\bibinfo {author} {\bibfnamefont {M.}~\bibnamefont
  {Mafu}}, \bibinfo {author} {\bibfnamefont {A.}~\bibnamefont {Dudley}},
  \bibinfo {author} {\bibfnamefont {S.}~\bibnamefont {Goyal}}, \bibinfo
  {author} {\bibfnamefont {D.}~\bibnamefont {Giovannini}}, \bibinfo {author}
  {\bibfnamefont {M.}~\bibnamefont {McLaren}}, \bibinfo {author} {\bibfnamefont
  {M.~J.}\ \bibnamefont {Padgett}}, \bibinfo {author} {\bibfnamefont
  {T.}~\bibnamefont {Konrad}}, \bibinfo {author} {\bibfnamefont
  {F.}~\bibnamefont {Petruccione}}, \bibinfo {author} {\bibfnamefont
  {N.}~\bibnamefont {L\"utkenhaus}}, \ and\ \bibinfo {author} {\bibfnamefont
  {A.}~\bibnamefont {Forbes}},\ }\href {\doibase 10.1103/PhysRevA.88.032305}
  {\bibfield  {journal} {\bibinfo  {journal} {Phys. Rev. A}\ }\textbf {\bibinfo
  {volume} {88}},\ \bibinfo {pages} {032305} (\bibinfo {year}
  {2013})}\BibitemShut {NoStop}%
\bibitem [{\citenamefont {Heo}\ \emph {et~al.}(2017)\citenamefont {Heo},
  \citenamefont {Kang}, \citenamefont {Hong}, \citenamefont {Yang},
  \citenamefont {Choi},\ and\ \citenamefont {Hong}}]{Heo2017}%
  \BibitemOpen
  \bibfield  {author} {\bibinfo {author} {\bibfnamefont {J.}~\bibnamefont
  {Heo}}, \bibinfo {author} {\bibfnamefont {M.-S.}\ \bibnamefont {Kang}},
  \bibinfo {author} {\bibfnamefont {C.-H.}\ \bibnamefont {Hong}}, \bibinfo
  {author} {\bibfnamefont {H.-J.}\ \bibnamefont {Yang}}, \bibinfo {author}
  {\bibfnamefont {S.-G.}\ \bibnamefont {Choi}}, \ and\ \bibinfo {author}
  {\bibfnamefont {J.-P.}\ \bibnamefont {Hong}},\ }\href {\doibase
  10.1038/s41598-017-09510-9} {\bibfield  {journal} {\bibinfo  {journal}
  {Scientific Reports}\ }\textbf {\bibinfo {volume} {7}},\ \bibinfo {pages}
  {10208} (\bibinfo {year} {2017})}\BibitemShut {NoStop}%
\bibitem [{\citenamefont {Erhard}\ \emph {et~al.}(2015)\citenamefont {Erhard},
  \citenamefont {Qassim}, \citenamefont {Mand}, \citenamefont {Karimi},\ and\
  \citenamefont {Boyd}}]{PhysRevA.92.022321}%
  \BibitemOpen
  \bibfield  {author} {\bibinfo {author} {\bibfnamefont {M.}~\bibnamefont
  {Erhard}}, \bibinfo {author} {\bibfnamefont {H.}~\bibnamefont {Qassim}},
  \bibinfo {author} {\bibfnamefont {H.}~\bibnamefont {Mand}}, \bibinfo {author}
  {\bibfnamefont {E.}~\bibnamefont {Karimi}}, \ and\ \bibinfo {author}
  {\bibfnamefont {R.~W.}\ \bibnamefont {Boyd}},\ }\href {\doibase
  10.1103/PhysRevA.92.022321} {\bibfield  {journal} {\bibinfo  {journal} {Phys.
  Rev. A}\ }\textbf {\bibinfo {volume} {92}},\ \bibinfo {pages} {022321}
  (\bibinfo {year} {2015})}\BibitemShut {NoStop}%
\bibitem [{\citenamefont {Fickler}\ \emph {et~al.}(2014)\citenamefont
  {Fickler}, \citenamefont {Lapkiewicz}, \citenamefont {Ramelow},\ and\
  \citenamefont {Zeilinger}}]{PhysRevA.89.060301}%
  \BibitemOpen
  \bibfield  {author} {\bibinfo {author} {\bibfnamefont {R.}~\bibnamefont
  {Fickler}}, \bibinfo {author} {\bibfnamefont {R.}~\bibnamefont {Lapkiewicz}},
  \bibinfo {author} {\bibfnamefont {S.}~\bibnamefont {Ramelow}}, \ and\
  \bibinfo {author} {\bibfnamefont {A.}~\bibnamefont {Zeilinger}},\ }\href
  {\doibase 10.1103/PhysRevA.89.060301} {\bibfield  {journal} {\bibinfo
  {journal} {Phys. Rev. A}\ }\textbf {\bibinfo {volume} {89}},\ \bibinfo
  {pages} {060301} (\bibinfo {year} {2014})}\BibitemShut {NoStop}%
\bibitem [{\citenamefont {Nape}\ \emph {et~al.}(2021)\citenamefont {Nape},
  \citenamefont {Mashaba}, \citenamefont {Mphuthi}, \citenamefont {Jayakumar},
  \citenamefont {Bhattacharya},\ and\ \citenamefont {Forbes}}]{turbulence1}%
  \BibitemOpen
  \bibfield  {author} {\bibinfo {author} {\bibfnamefont {I.}~\bibnamefont
  {Nape}}, \bibinfo {author} {\bibfnamefont {N.}~\bibnamefont {Mashaba}},
  \bibinfo {author} {\bibfnamefont {N.}~\bibnamefont {Mphuthi}}, \bibinfo
  {author} {\bibfnamefont {S.}~\bibnamefont {Jayakumar}}, \bibinfo {author}
  {\bibfnamefont {S.}~\bibnamefont {Bhattacharya}}, \ and\ \bibinfo {author}
  {\bibfnamefont {A.}~\bibnamefont {Forbes}},\ }\href {\doibase
  10.1103/PhysRevApplied.15.034030} {\bibfield  {journal} {\bibinfo  {journal}
  {Phys. Rev. Applied}\ }\textbf {\bibinfo {volume} {15}},\ \bibinfo {pages}
  {034030} (\bibinfo {year} {2021})}\BibitemShut {NoStop}%
\bibitem [{\citenamefont {Zhou}\ \emph {et~al.}(2021)\citenamefont {Zhou},
  \citenamefont {Zhao}, \citenamefont {Braverman}, \citenamefont {Pang},
  \citenamefont {Zhang}, \citenamefont {Willner}, \citenamefont {Shi},\ and\
  \citenamefont {Boyd}}]{turbulence3}%
  \BibitemOpen
  \bibfield  {author} {\bibinfo {author} {\bibfnamefont {Y.}~\bibnamefont
  {Zhou}}, \bibinfo {author} {\bibfnamefont {J.}~\bibnamefont {Zhao}}, \bibinfo
  {author} {\bibfnamefont {B.}~\bibnamefont {Braverman}}, \bibinfo {author}
  {\bibfnamefont {K.}~\bibnamefont {Pang}}, \bibinfo {author} {\bibfnamefont
  {R.}~\bibnamefont {Zhang}}, \bibinfo {author} {\bibfnamefont {A.~E.}\
  \bibnamefont {Willner}}, \bibinfo {author} {\bibfnamefont {Z.}~\bibnamefont
  {Shi}}, \ and\ \bibinfo {author} {\bibfnamefont {R.~W.}\ \bibnamefont
  {Boyd}},\ }\href {\doibase 10.1103/PhysRevApplied.15.034011} {\bibfield
  {journal} {\bibinfo  {journal} {Phys. Rev. Applied}\ }\textbf {\bibinfo
  {volume} {15}},\ \bibinfo {pages} {034011} (\bibinfo {year}
  {2021})}\BibitemShut {NoStop}%
\bibitem [{\citenamefont {Gianani}\ \emph {et~al.}(2020)\citenamefont
  {Gianani}, \citenamefont {Suprano}, \citenamefont {Giordani}, \citenamefont
  {Spagnolo}, \citenamefont {Sciarrino}, \citenamefont {Gorpas}, \citenamefont
  {Ntziachristos}, \citenamefont {Pinker}, \citenamefont {Biton}, \citenamefont
  {Kupferman},\ and\ \citenamefont {Arnon}}]{turbulence2}%
  \BibitemOpen
  \bibfield  {author} {\bibinfo {author} {\bibfnamefont {I.}~\bibnamefont
  {Gianani}}, \bibinfo {author} {\bibfnamefont {A.}~\bibnamefont {Suprano}},
  \bibinfo {author} {\bibfnamefont {T.}~\bibnamefont {Giordani}}, \bibinfo
  {author} {\bibfnamefont {N.}~\bibnamefont {Spagnolo}}, \bibinfo {author}
  {\bibfnamefont {F.}~\bibnamefont {Sciarrino}}, \bibinfo {author}
  {\bibfnamefont {D.}~\bibnamefont {Gorpas}}, \bibinfo {author} {\bibfnamefont
  {V.}~\bibnamefont {Ntziachristos}}, \bibinfo {author} {\bibfnamefont
  {K.}~\bibnamefont {Pinker}}, \bibinfo {author} {\bibfnamefont
  {N.}~\bibnamefont {Biton}}, \bibinfo {author} {\bibfnamefont
  {J.}~\bibnamefont {Kupferman}}, \ and\ \bibinfo {author} {\bibfnamefont
  {S.}~\bibnamefont {Arnon}},\ }\href {\doibase 10.1117/1.AP.2.3.036003}
  {\bibfield  {journal} {\bibinfo  {journal} {Advanced Photonics}\ }\textbf
  {\bibinfo {volume} {2}},\ \bibinfo {pages} {1 } (\bibinfo {year}
  {2020})}\BibitemShut {NoStop}%
\bibitem [{\citenamefont {Pati}(2000)}]{Pati2000}%
  \BibitemOpen
  \bibfield  {author} {\bibinfo {author} {\bibfnamefont {A.~K.}\ \bibnamefont
  {Pati}},\ }\href {\doibase 10.1103/PhysRevA.63.014302} {\bibfield  {journal}
  {\bibinfo  {journal} {Phys. Rev. A}\ }\textbf {\bibinfo {volume} {63}},\
  \bibinfo {pages} {014302} (\bibinfo {year} {2000})}\BibitemShut {NoStop}%
\bibitem [{\citenamefont {Bennett}\ \emph {et~al.}(2001)\citenamefont
  {Bennett}, \citenamefont {DiVincenzo}, \citenamefont {Shor}, \citenamefont
  {Smolin}, \citenamefont {Terhal},\ and\ \citenamefont
  {Wootters}}]{Bennett2001}%
  \BibitemOpen
  \bibfield  {author} {\bibinfo {author} {\bibfnamefont {C.~H.}\ \bibnamefont
  {Bennett}}, \bibinfo {author} {\bibfnamefont {D.~P.}\ \bibnamefont
  {DiVincenzo}}, \bibinfo {author} {\bibfnamefont {P.~W.}\ \bibnamefont
  {Shor}}, \bibinfo {author} {\bibfnamefont {J.~A.}\ \bibnamefont {Smolin}},
  \bibinfo {author} {\bibfnamefont {B.~M.}\ \bibnamefont {Terhal}}, \ and\
  \bibinfo {author} {\bibfnamefont {W.~K.}\ \bibnamefont {Wootters}},\ }\href
  {\doibase 10.1103/PhysRevLett.87.077902} {\bibfield  {journal} {\bibinfo
  {journal} {Phys. Rev. Lett.}\ }\textbf {\bibinfo {volume} {87}},\ \bibinfo
  {pages} {077902} (\bibinfo {year} {2001})}\BibitemShut {NoStop}%
\bibitem [{\citenamefont {Lo}(2000)}]{Lo2000}%
  \BibitemOpen
  \bibfield  {author} {\bibinfo {author} {\bibfnamefont {H.-K.}\ \bibnamefont
  {Lo}},\ }\href {\doibase 10.1103/PhysRevA.62.012313} {\bibfield  {journal}
  {\bibinfo  {journal} {Phys. Rev. A}\ }\textbf {\bibinfo {volume} {62}},\
  \bibinfo {pages} {012313} (\bibinfo {year} {2000})}\BibitemShut {NoStop}%
\bibitem [{\citenamefont {Leung}\ and\ \citenamefont {Shor}(2003)}]{Leung2003}%
  \BibitemOpen
  \bibfield  {author} {\bibinfo {author} {\bibfnamefont {D.~W.}\ \bibnamefont
  {Leung}}\ and\ \bibinfo {author} {\bibfnamefont {P.~W.}\ \bibnamefont
  {Shor}},\ }\href {\doibase 10.1103/PhysRevLett.90.127905} {\bibfield
  {journal} {\bibinfo  {journal} {Phys. Rev. Lett.}\ }\textbf {\bibinfo
  {volume} {90}},\ \bibinfo {pages} {127905} (\bibinfo {year}
  {2003})}\BibitemShut {NoStop}%
\bibitem [{\citenamefont {Daki{\'{c}}}\ \emph {et~al.}(2012)\citenamefont
  {Daki{\'{c}}}, \citenamefont {Lipp}, \citenamefont {Ma}, \citenamefont
  {Ringbauer}, \citenamefont {Kropatschek}, \citenamefont {Barz}, \citenamefont
  {Paterek}, \citenamefont {Vedral}, \citenamefont {Zeilinger}, \citenamefont
  {Brukner},\ and\ \citenamefont {Walther}}]{Dakic2012}%
  \BibitemOpen
  \bibfield  {author} {\bibinfo {author} {\bibfnamefont {B.}~\bibnamefont
  {Daki{\'{c}}}}, \bibinfo {author} {\bibfnamefont {Y.~O.}\ \bibnamefont
  {Lipp}}, \bibinfo {author} {\bibfnamefont {X.}~\bibnamefont {Ma}}, \bibinfo
  {author} {\bibfnamefont {M.}~\bibnamefont {Ringbauer}}, \bibinfo {author}
  {\bibfnamefont {S.}~\bibnamefont {Kropatschek}}, \bibinfo {author}
  {\bibfnamefont {S.}~\bibnamefont {Barz}}, \bibinfo {author} {\bibfnamefont
  {T.}~\bibnamefont {Paterek}}, \bibinfo {author} {\bibfnamefont
  {V.}~\bibnamefont {Vedral}}, \bibinfo {author} {\bibfnamefont
  {A.}~\bibnamefont {Zeilinger}}, \bibinfo {author} {\bibfnamefont
  {{\v{C}}.}~\bibnamefont {Brukner}}, \ and\ \bibinfo {author} {\bibfnamefont
  {P.}~\bibnamefont {Walther}},\ }\href {\doibase 10.1038/nphys2377} {\bibfield
   {journal} {\bibinfo  {journal} {Nature Physics}\ }\textbf {\bibinfo {volume}
  {8}},\ \bibinfo {pages} {666} (\bibinfo {year} {2012})}\BibitemShut {NoStop}%
\bibitem [{\citenamefont {Barreiro}\ \emph {et~al.}(2010)\citenamefont
  {Barreiro}, \citenamefont {Wei},\ and\ \citenamefont
  {Kwiat}}]{barreiro2010remote}%
  \BibitemOpen
  \bibfield  {author} {\bibinfo {author} {\bibfnamefont {J.~T.}\ \bibnamefont
  {Barreiro}}, \bibinfo {author} {\bibfnamefont {T.-C.}\ \bibnamefont {Wei}}, \
  and\ \bibinfo {author} {\bibfnamefont {P.~G.}\ \bibnamefont {Kwiat}},\ }\href
  {\doibase 10.1103/PhysRevLett.105.030407} {\bibfield  {journal} {\bibinfo
  {journal} {Phys. Rev. Lett.}\ }\textbf {\bibinfo {volume} {105}},\ \bibinfo
  {pages} {030407} (\bibinfo {year} {2010})}\BibitemShut {NoStop}%
\bibitem [{\citenamefont {Peters}\ \emph {et~al.}(2005)\citenamefont {Peters},
  \citenamefont {Barreiro}, \citenamefont {Goggin}, \citenamefont {Wei},\ and\
  \citenamefont {Kwiat}}]{Peters2005}%
  \BibitemOpen
  \bibfield  {author} {\bibinfo {author} {\bibfnamefont {N.~A.}\ \bibnamefont
  {Peters}}, \bibinfo {author} {\bibfnamefont {J.~T.}\ \bibnamefont
  {Barreiro}}, \bibinfo {author} {\bibfnamefont {M.~E.}\ \bibnamefont
  {Goggin}}, \bibinfo {author} {\bibfnamefont {T.-C.}\ \bibnamefont {Wei}}, \
  and\ \bibinfo {author} {\bibfnamefont {P.~G.}\ \bibnamefont {Kwiat}},\ }\href
  {\doibase 10.1103/PhysRevLett.94.150502} {\bibfield  {journal} {\bibinfo
  {journal} {Phys. Rev. Lett.}\ }\textbf {\bibinfo {volume} {94}},\ \bibinfo
  {pages} {150502} (\bibinfo {year} {2005})}\BibitemShut {NoStop}%
\bibitem [{\citenamefont {Sarenac}\ \emph
  {et~al.}(2018{\natexlab{b}})\citenamefont {Sarenac}, \citenamefont {Cory},
  \citenamefont {Nsofini}, \citenamefont {Hincks}, \citenamefont {Miguel},
  \citenamefont {Arif}, \citenamefont {Clark}, \citenamefont {Huber},\ and\
  \citenamefont {Pushin}}]{sarenac2018generation}%
  \BibitemOpen
  \bibfield  {author} {\bibinfo {author} {\bibfnamefont {D.}~\bibnamefont
  {Sarenac}}, \bibinfo {author} {\bibfnamefont {D.~G.}\ \bibnamefont {Cory}},
  \bibinfo {author} {\bibfnamefont {J.}~\bibnamefont {Nsofini}}, \bibinfo
  {author} {\bibfnamefont {I.}~\bibnamefont {Hincks}}, \bibinfo {author}
  {\bibfnamefont {P.}~\bibnamefont {Miguel}}, \bibinfo {author} {\bibfnamefont
  {M.}~\bibnamefont {Arif}}, \bibinfo {author} {\bibfnamefont {C.~W.}\
  \bibnamefont {Clark}}, \bibinfo {author} {\bibfnamefont {M.~G.}\ \bibnamefont
  {Huber}}, \ and\ \bibinfo {author} {\bibfnamefont {D.~A.}\ \bibnamefont
  {Pushin}},\ }\href {\doibase 10.1103/PhysRevLett.121.183602} {\bibfield
  {journal} {\bibinfo  {journal} {Phys. Rev. Lett.}\ }\textbf {\bibinfo
  {volume} {121}},\ \bibinfo {pages} {183602} (\bibinfo {year}
  {2018}{\natexlab{b}})}\BibitemShut {NoStop}%
\bibitem [{\citenamefont {James}\ \emph {et~al.}(2001)\citenamefont {James},
  \citenamefont {Kwiat}, \citenamefont {Munro},\ and\ \citenamefont
  {White}}]{James2001}%
  \BibitemOpen
  \bibfield  {author} {\bibinfo {author} {\bibfnamefont {D.~F.~V.}\
  \bibnamefont {James}}, \bibinfo {author} {\bibfnamefont {P.~G.}\ \bibnamefont
  {Kwiat}}, \bibinfo {author} {\bibfnamefont {W.~J.}\ \bibnamefont {Munro}}, \
  and\ \bibinfo {author} {\bibfnamefont {A.~G.}\ \bibnamefont {White}},\ }\href
  {\doibase 10.1103/PhysRevA.64.052312} {\bibfield  {journal} {\bibinfo
  {journal} {Phys. Rev. A}\ }\textbf {\bibinfo {volume} {64}},\ \bibinfo
  {pages} {052312} (\bibinfo {year} {2001})}\BibitemShut {NoStop}%
\bibitem [{\citenamefont {Horodecki}\ \emph {et~al.}(2009)\citenamefont
  {Horodecki}, \citenamefont {Horodecki}, \citenamefont {Horodecki},\ and\
  \citenamefont {Horodecki}}]{RevModPhys.81.865}%
  \BibitemOpen
  \bibfield  {author} {\bibinfo {author} {\bibfnamefont {R.}~\bibnamefont
  {Horodecki}}, \bibinfo {author} {\bibfnamefont {P.}~\bibnamefont
  {Horodecki}}, \bibinfo {author} {\bibfnamefont {M.}~\bibnamefont
  {Horodecki}}, \ and\ \bibinfo {author} {\bibfnamefont {K.}~\bibnamefont
  {Horodecki}},\ }\href {\doibase 10.1103/RevModPhys.81.865} {\bibfield
  {journal} {\bibinfo  {journal} {Rev. Mod. Phys.}\ }\textbf {\bibinfo {volume}
  {81}},\ \bibinfo {pages} {865} (\bibinfo {year} {2009})}\BibitemShut
  {NoStop}%
\end{thebibliography}%

\end{document}